\documentclass[amssymb,prl,twocolumn,showpacs,reprint]{revtex4}
\usepackage{amsmath}
\usepackage{bbold}
\usepackage{amsbsy}
\usepackage{graphicx}
\usepackage{overpic}
\usepackage{verbatim}
\usepackage{graphics}

\newcommand{\bea}{\begin{eqnarray}}
\newcommand{\eea}{\end{eqnarray}}
\newcommand{\be}{\begin{equation}}
\newcommand{\ee}{\end{equation}}

\begin{document}

\title{
Dark Bell states in tunnel-coupled spin qubits}
\author{Rafael S\'anchez}
\affiliation{Instituto de Ciencia de Materiales de Madrid (ICMM-CSIC), Cantoblanco 28049 Madrid, Spain}
\author{Gloria Platero}
\affiliation{Instituto de Ciencia de Materiales de Madrid (ICMM-CSIC), Cantoblanco 28049 Madrid, Spain}
\date{\today}

\begin{abstract}
We investigate the dynamical purification of maximally entangled electron states by transport through coupled quantum dots. Under resonant ac driving and coherent tunneling, even-parity Bell states perform Rabi oscillations that decouple from the environment, leading to a dark state. The two electrons remain spatially separated, one in each quantum dot. We propose configurations where this effect will prove as antiresonances in transport spectroscopy experiments.
\end{abstract}
%#73.23.Hk 
%Coulomb blockade; single-electron tunneling
%# 73.23.-b
%Electronic transport in mesoscopic systems
%#03.67.Bg
%Entanglement production and manipulation
%# 72.70.+m
%Noise processes and phenomena in electronic transport
%# 73.63.Kv
%Quantum dots (electronic transport) 
\maketitle

%\section{Introduction}
{\it Introduction---}
Quantum coherence allows open multi-level systems to form superpositions that uncouple from the dissipative dynamics. The system evolves towards a stationary pure state. These are named dark states from their discovery as non-absorbing resonances in illuminated atom gases~\cite{cpt}. Two ground states coupled to the same excited state form a superposition that is not affected by the lasers. The occupation of orthogonal superpositions which can be excited by the lasers (often called bright states) decays due to the spontaneous emission from the excited state. Many applications for optical systems followed, including laser cooling, lasing without inversion and coherent adiabatic passage~\cite{arimondo}. 

As for some remarkable quantum optics phenomena, these applications have found their translation to mesoscopic electronic circuits~\cite{tobias}. 
There, transitions are mediated by coherent tunneling or by time dependent electromagnetic fields. In quantum dot arrays, dark superpositions are essential to proposals of coherent state transfer~\cite{greentree} or current switching by coherent population trapping~\cite{tobiascpt}. However, their  detection in transport has been elusive for requiring exquisite control of complicated multidot arrangements~\cite{michaelis,maria} or many level configurations~\cite{clive}. Here we introduce a dark state based on collective spin dynamics rather than single-electron interference. The required technology is the same that has achieved single electron spin resonance (ESR)~\cite{engel2,koppens} in already a number of experiments~\cite{koppens,laird,nowack,pioro,nadjperge}. Remarkably, the resulting two-electron dark state is maximally entangled.

%Importantly for this work, single electron spin resonance (ESR) has been achieved in quantum dot spin qubits driven by ac and dc magnetic fields~\cite{engel2,koppens}. Here we propose that the same setup can be used to detect a two electron dark state.

%Of great interest is the generation of entangled states. Entangling two qubits enables quantum information protocols. 
In the solid state, entanglement has been demonstrated by the violation of Bell inequalities in superconductor phase qubits~\cite{ansmann} while quantum dot spin qubits have succeeded in the performance of essential ingredients as universal quantum gates~\cite{petta,koppens} or single shot readout~\cite{nowack-2qubit}. Electronic entanglers have been proposed based on Cooper pair splitting~\cite{recher}, electron-hole excitations~\cite{beenakker}, single-electron emitters~\cite{janine}, parity detection~\cite{trauzettel} or purification protocols~\cite{taylor}. Quite counter-intuitively, entanglement of macroscopic atomic ensembles can be generated by dissipation~\cite{muschik}. A proper engineering of the environment drives an open quantum system to a pure steady state~\cite{diehl}, similar to a dark state. 

\begin{figure}[b]
\centering
%\includegraphics[width=0.34\linewidth,trim=0 -20 0 -20,clip=true]{paritysubsp.eps}
%\vspace{0.2cm}
%\includegraphics[width=0.64\linewidth]{pijtime.eps}
%\includegraphics[width=\linewidth,clip]{purif.eps}
\includegraphics[width=0.9\linewidth,clip]{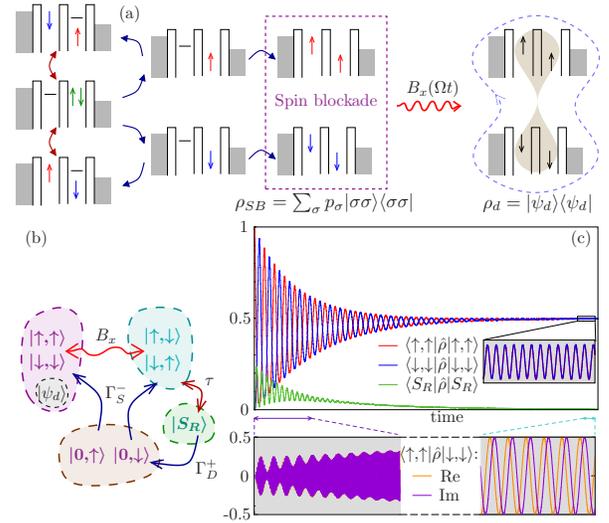}
\caption{\label{sys} Dynamical generation of entanglement. (a) In a spin blockade double quantum dot, charge flows through states of two electrons with opposite spins, $\{{\mid}{\uparrow}{,}{\downarrow}\rangle,{\mid}{\downarrow}{,}{\uparrow}\rangle\}$, coupled by intradot tunneling, $\tau$, to the doubly occupied singlet ${\mid}S_R\rangle$, from where an electron tunnels to the drain lead. (b) The occupation of the even-parity subspace blocks the current, forming a mixed state, $\hat\rho_{SB}$, with statistical weights, $p_\sigma$. Coherent spin rotations by an ac magnetic field, $B_x$, open the system to transport again. For a particular frequency, $\Omega$, an even-parity eigenstate of the magnetic field becomes a dark state, ${\mid}\psi_d\rangle$. (c) The system evolves towards a pure steady state composed of Bell superpositions. Coherence of the steady state is apparent in the undamped oscillations of the off-diagonal elements.
}
\end{figure}

In solid state qubits, the environment can be engineered by voltages and electromagnetic fields.
Here we propose how to generate a maximally entangled dark state 
%Our dark state consists on a maximally entangled superposition
of two spatially separated electrons stored in a double quantum dot tunnel-coupled in series to fermionic source and drain leads. In the Coulomb blockade regime with strong Coulomb interactions, electrons are transferred one by one through the system, which can be tuned to contain up to two conduction electrons, cf. Fig.~\ref{sys}. We consider a configuration described by the charge occupation states ${(}N_L{,}N_R{)}{=}{(}0{,}1{)}{,}{(}1{,}1{)}{,}{(}0{,}2{)}$. 
In the presence of inhomogeneous in-plane magnetic fields, odd-parity states, $\{{\mid}{\uparrow}{,}{\downarrow}\rangle,{\mid}{\downarrow}{,}{\uparrow}\rangle\}$, support a current through the drain dot singlet ${\mid}S_R\rangle{=}{\mid}{0}{,}{\uparrow}{\downarrow}\rangle$ from which an electron is transferred to the drain lead. However, the occupation of the even parity subspace, $\{{\mid}{\uparrow}{,}{\uparrow}\rangle,{\mid}{\downarrow}{,}{\downarrow}\rangle\}$, suppresses the current for a forward applied bias (from left to right) due to Pauli exclusion principle, what is known as spin blockade~\cite{ono}.
%, where spin-spin correlations are enhanced~\cite{spin}. 
Thus, the reduced density matrix of the double quantum dot, $\hat\rho$, evolves toward a mixed steady state  in the even parity subspace: $\hat\rho_{SB}{=}p_\uparrow{\mid}{\uparrow}{,}{\uparrow}\rangle\langle{\uparrow}{,}{\uparrow}{\mid}{+}p_\downarrow{\mid}{\downarrow}{,}{\downarrow}\rangle\langle{\downarrow}{,}{\downarrow}{\mid}$, with statistical weights $p_\sigma$. Single electron spin rotations by  an ac magnetic field lift spin blockade and lead to a resonant current when its frequency matches one of the individual Zeeman splittings, $\hbar\omega{=}\Delta_L,\Delta_R$~\cite{koppens,pioro,nadjperge}. We find that for a particular frequency, superpositions of even-parity Bell states, 
\be
{\mid}\psi_\pm\rangle{=}\frac{1}{\sqrt{2}}({\mid}{\uparrow}{,}{\uparrow}\rangle{\pm}{\mid}{\downarrow}{,}{\downarrow}\rangle),
\ee 
decouple both from the driving field and transport. The system is thus dynamically driven to a pure dark state, $\rho_{st}{\approx}{\mid}\psi_d\rangle\langle\psi_d{\mid}$, see Eqs.~\eqref{ds} and \eqref{rst} below. Any other superposition decays by the combined effect of the ac field and tunneling into the reservoirs. 

We investigate the fingerprints that such states leave in transport spectroscopy experiments as being carried out nowadays~\cite{koppens,laird,nowack,pioro,nadjperge}. Collective rotations of the two electron spins lead to a resonant current showing a sharp dip which cannot be explained by the individual electron spin dynamics. This effect will manifest clearly  in the current at the border of the spin blockade window where the (1,1)$\rightarrow$(0,2) tunneling transition is resonant. 

%\section{Model}
{\it Model---}
Our system is modeled by a two-site Anderson Hamiltonian,
$\hat H{=}\hat H_{\rm DQD}{+}\hat H_{\rm leads}{+}\hat H_{\rm coupl}$, describing the double quantum dot, the electronic source ($S$) and drain ($D$) reservoirs, $\hat{H}_{\rm leads}{=}\sum_{lk\sigma}\varepsilon_{lk}\hat d_{lk\sigma}^\dagger\hat d_{lk\sigma}$, and their tunneling couplings, $\hat H_{\rm coupl}{=}{\sum_{lk\sigma}}(\Lambda_{l}\hat{d}_{lk\sigma}^\dagger\hat{c}_{l\sigma}{+}{\rm H.c.})$,
where the fermionic operators $\hat{c}_{i\sigma}$ and $\hat{d}_{lk\sigma}$ annihilate an electron with spin $\sigma$ in dot $i{\in}\{L,R\}$ and lead $l{\in}\{S,D\}$, respectively, and $\Lambda_l$ is the coupling strength. The double quantum dot term is given by $\hat{H}_{\rm DQD}{=}\hat{H}_0{+}\hat{H}_\tau$, accounting for the bare energy of the double quantum dot (including interactions), $\hat{H}_0$, and the coherent interdot tunneling, $\hat{H}_\tau{=}{-}\sum_{\sigma}(\tau\hat{c}_{\rm L\sigma}^\dagger\hat{c}_{\rm R\sigma}{+}{\rm H.c.})$. We consider a single discrete level of energy $\varepsilon_i$ in each dot which can be occupied by two electrons forming a spin singlet. Excited states are assumed to be far off in energy so their contribution ---forming on-site triplets that would lift spin blockade--- can be neglected here. Thus we have
$\hat{H}_{0}=\sum_{i\sigma}\varepsilon_{i}\hat c_{i\sigma}^\dagger\hat c_{i\sigma}+\sum_{i}U_{i}\hat n_{i{{\uparrow}}}\hat n_{i{\downarrow}}+U_{LR}\hat n_{L}\hat n_{R}$,
where $U_i$ and $U_{LR}$ describe on-site and interdot Coulomb repulsion, and $\hat{n}_{i\sigma}$ and $\hat{n}_{i}$ are the spin resolved and total number operators, respectively.
The chemical potentials of the leads, $\mu_i$, are such that only two electrons are
allowed in the system: 
$\varepsilon_i{<}\mu_i{-}U_{\rm LR}{<}\varepsilon_i{+}U_i$ and $\mu_i{<}\varepsilon_i{+}2U_{LR}$. If $\varepsilon_R{<}\mu_D$, the right quantum dot stays occupied. Considering spin, only seven states with charge distributions described by (0,1), (1,1) and (0,2) are relevant. 

Coherent interdot tunneling will be resonant when the states (1,1) and (0,2) are degenerate. We parametrize their detuning by $\delta\varepsilon{=}\varepsilon_L{-}\varepsilon_R{+}U_R{-}U_{LR}$. The spin blockade window is hence defined by the region $\delta\varepsilon\ge0$.

For spin manipulation, we include magnetic fields in an ESR configuration~\cite{engel2,moi}. It consists of in-plane magnetic fields with a dc component that creates a Zeeman splitting $\Delta^z_i$ in each dot, and an ac component perpendicular to it whose frequency is close to the resonance conditions $\hbar\omega{\approx}\Delta^z_i$. Since magnetic fields are experimentally hard to localize, alternatives have been introduced using more tunable gate voltages. Effective ac magnetic fields are obtained by coupling the spin to a pulsed electric signal mediated by either spin orbit or hyperfine interaction~\cite{nowack,nadjperge,laird}, slanting dc magnetic fields~\cite{tokura,pioro}, or by spin-phonon coupling in the presence of mechanical vibrations of the dot~\cite{ohm}. For theoretical generality, we consider the purely magnetic field case given by the Hamiltonian term:
$\hat{H}_B(t){=}\sum_i(\Delta^x_i\cos\omega t,0,\Delta^z_i){\cdot}\mathbf{\hat{S}_i}$,
with $\Delta^j_i{=}g_{ji}\mu_BB_{ji}$ and spin operators ${\bf \hat S}_i{=}\frac{1}{2}\sum_{\sigma \sigma'} \hat c^\dagger_{i\sigma} \boldsymbol{\hat\sigma}_{\sigma \sigma'}\hat c_{i\sigma'}$. The Land\'e factors, $g_{ji}$, are in general inhomogeneous, $\mu_B$ is the Bohr magneton, $B_{ji}$ are the magnetic field $j$ component in dot $i$, and $ \boldsymbol{\hat\sigma}$ are the Pauli matrices. Different Zeeman splittings are required, $\Delta^z_L{\ne}\Delta^z_R$, otherwise the ESR field has no visible effect~\cite{moi}. The different components $\Delta^j_{i}$ can be tuned either by the control of the Land\'e factor~\cite{huang} or by applying inhomogeneous magnetic fields: for instance by means of a micromagnet~\cite{pioro,laird}, or pulses of different amplitude in electric dipole spin resonance experiments. Such experimental ability allows us to consider the simplest required configuration, with $\Delta^x_L{=}\Delta^x_R{=}\gamma B_x$ and $\Delta^z_L{=}a\Delta^z_R{=}\gamma B_z$. To ease the notation, we include the asymmetries in the parameter $a$, assuming that the magnetic fields are homogeneous.

The dynamical evolution of the system is described by a Markovian quantum master equation for the reduced density matrix $\hat\rho$ obtained by tracing the reservoir degrees of freedom out:%~\cite{stoof,gurvitz}:
%In the Born-Markov approximation, it reads~\cite{stoof,gurvitz}:
\be
\label{mastereq}
\dot\rho=-\frac{i}{\hbar}[\hat{H}_{\rm DQD}{+}\hat{H}_B(t),\hat\rho]+{\cal L}_\Gamma\hat\rho, 
\ee
where the commutator accounts for the coherent dynamics inside the double quantum dot. 
The Liouvillian superoperator ${\cal L}_\Gamma{=}\sum_{\alpha{=}\pm,l}{\cal L}_l^\alpha$ describes tunneling events to ($+$)  or from ($-$) lead $l$ and includes decoherence for the finite lifetime of states coupled to the leads~\cite{engel2,moi}. Spin decoherence is assumed to be of a longer time scale. The drain current is then given by $I{=}q{\rm tr}[({\cal L}_D^+-{\cal L}_D^-)\hat\rho]$. In the high bias regime, transport is unidirectional with ${\cal L}_S^+{=}{\cal L}_D^-{=}0$. The remaining terms are fully described by the tunneling rates $\Gamma_l{=}\frac{2\pi}{\hbar}|\Lambda_l|^2\nu_l$, where $\nu_l$ is the density of states in the leads, for processes carrying an electron from the source to the dots and from the dots to the drain. Processes taking electrons in the backward direction are negligible, though they are included in the numerical calculations. 
%The Liouvillian superoperator in the second term describes tunneling processes through the leads, including decoherence for the finite lifetime of states coupled to transport~\cite{engel2,moi}. Spin decoherence is assumed to be of a longer time scale. In the high bias regime, ${\cal L}_\Gamma$ is fully described by the tunneling rates $\Gamma_l{=}\frac{2\pi}{\hbar}|\Lambda_l|^2\nu_l$, where $\nu_l$ is the density of states in the leads, for processes carrying an electron from the source to the dots and from the dots to the drain. Processes taking electrons in the backwards direction are negligible, though they are included in the numerical calculations. 

\begin{figure}
\begin{center}
\includegraphics[width=0.685\linewidth,clip]{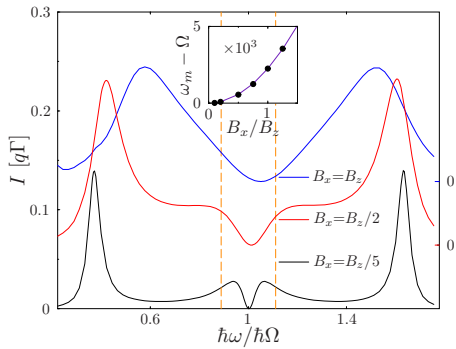}
\end{center}
\caption{\label{iwbac}\small Current as a function of the driving frequency, $\omega$. The collective spin rotation results in a nontrivial lineshape not centered at the ESR conditions (at $\hbar\omega{=}\Delta_L^z,\Delta_R^z$, vertical dashed lines) and a vanishing current (dark state) at $\hbar\Omega{=}(\Delta_L^z{+}\Delta_R^z)/2$. Cases for different driving amplitudes, $B_x$, are vertically offset. 
%Dashed lines mark the single ESR conditions. 
Parameters (for all figures, except where indicated): $\hbar\Gamma{=}10^{-2}$meV, $\tau{=}20\Gamma$, $\gamma{=}0.4$meV/T, $B_z{=}0.1$T and $a{=}0.8$. Interdot tunneling is resonant: $\delta\varepsilon{=}0$. The inset shows the shift of the antiresonance frequency, $\omega_m$, with the amplitude of the ac field. The solid line is a quadratic fit.}
\end{figure}

%\section{Results}
{\it Results---}
Na\"ively, one would expect that the current spectroscopy results from the sum of two Lorentzian resonance peaks centered at $\hbar\omega{=}\Delta_L^z,\Delta_R^z$ ---when the rotation of each electron spin would lift spin blockade--- and is zero elsewhere. Indeed, this picture agrees with experimental observations~\cite{pioro,nadjperge,laird}. However, due to the interplay of coherent spin rotation and interdot tunneling, resonant features of a non trivial lineshape appear, as shown in Fig.~\ref{iwbac}. On one hand, peaks are not necessarily centered at the individual spin resonances but at points of maximal hybridization which depend on $\tau$, $\Delta_i^j$ and $\delta\varepsilon$. On the other hand, and most importantly, current vanishes for a frequency $\hbar\omega{=}\hbar\Omega{\equiv}(\Delta_L^z{+}\Delta_R^z)/2$. Such anti-resonant behaviour is a clear signature of dark states. We note that similar current characteristics are predicted in discrete lattice models of transport~\cite{alfredo} to which our system can be mapped.

%\subsection{Rotating wave approximation}
Analytical understanding of the dynamics at that point can be obtained within a rotating wave approximation (RWA), when $\Delta_i^x{\ll}\Delta_i^z{\sim}\hbar\omega$. Thus, neglecting the contribution of counter-rotating terms, we get a time independent magnetic field Hamiltonian:
\be
\hat{H}_{B,{\rm RWA}}=\sum_i\left[(\Delta_i^z-\hbar\omega)\hat{S}_i^z+\Delta_i^x\hat{S}_i^x\right].
\ee
We can easily verify that ${\mid}\psi_-\rangle{=}({\mid}{\uparrow}{\uparrow}\rangle{-}{\mid}{\downarrow}{\downarrow}\rangle)/\sqrt{2}$ is a zero eigenvalue eigenstate of $\hat{H}_{B,{\rm RWA}}+\hat{H}_{\tau}$ at $\omega{=}\Omega$. Therefore it is decoupled both from the external magnetic field and from transport, due to parity and spin blockade. Electrons occupying any other state or superposition will flow to the drain lead and be replaced. As a consequence, the system will dynamically evolve towards a stationary solution given by the Bell state ${\mid}\psi_-\rangle$. The density matrix will be then described by the pure state ${\mid}\psi_-\rangle\langle\psi_-{\mid}$, fulfilling ${\cal L}_{\rm RWA}{\mid}\psi_-\rangle\langle\psi_-{\mid}{=}0$, with the Liouvillian superoperator ${\cal L}_{\rm RWA}\hat{\cal O}{=}{-}i\hbar^{-1}[\hat{H}_{\rm DQD}{+}\hat{H}_{B,{\rm RWA}},\hat{\cal O}]{+}{\cal L}_\Gamma\hat{\cal O}$. In this sense, opening the system to transport drives it to a maximally entangled dark state, for any initial state. 

In the laboratory frame, the steady state describes Rabi oscillations of the two even-parity Bell states, ${\mid}\psi_\pm\rangle$, with frequency $\Omega{=}(\Delta_L^z{+}\Delta_R^z)/(2\hbar)$:
\be
\label{ds}
{\mid}\psi_{st}(t)\rangle{=}i\sin\Omega t{\mid}\psi_+\rangle{+}\cos\Omega t{\mid}\psi_-\rangle,
\ee
within the RWA.
Hence we can approximate the stationary density matrix by ${\mid}\psi_{st}(t)\rangle\langle\psi_{st}(t){\mid}$, explicitly:
\be
\label{rst}
\hat\rho_{st}(t){\approx}{\frac{1}{2}}{\left[{\mid}{\uparrow}{,}{\uparrow}\rangle\langle{\uparrow}{,}{\uparrow}{\mid}{+}{\mid}{\downarrow}{,}{\downarrow}\rangle\langle {\downarrow}{,}{\downarrow}{\mid}{-}{\left(e^{2i\Omega t}{\mid}{\uparrow}{,}{\uparrow}\rangle\langle {\downarrow}{,}{\downarrow}{\mid}{+}{\rm H.c.}\right)}\right]}{.}
\ee
We note that this solution is exact for a circularly polarized ac magnetic field but, for experimental feasibility, we restrict our analysis to linear polarization. As known since the work of Bloch and Siegert~\cite{blochsiegert}, counter-rotating terms of linearly polarized electromagnetic fields induce a shift of the resonance condition; see inset in Fig.~\ref{iwbac}.

\begin{figure}
%\begin{center}
%\flushleft{
\includegraphics[width=0.3\linewidth,clip]{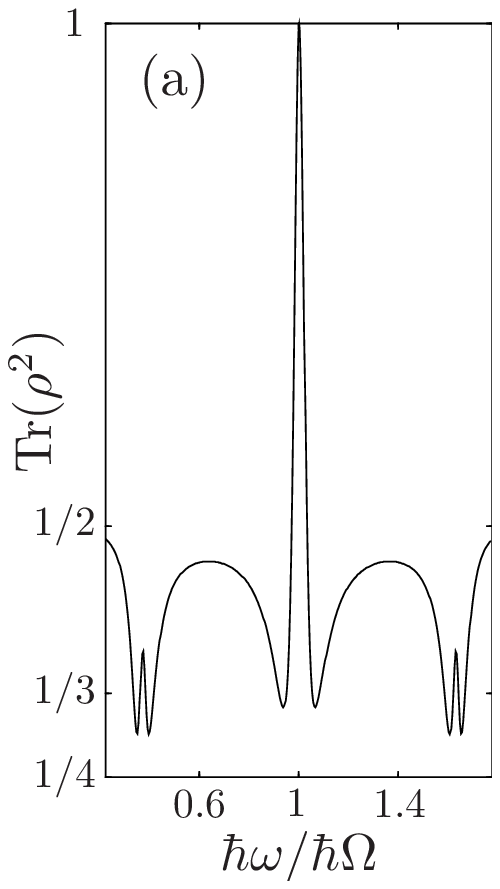}
%}
\includegraphics[width=0.685\linewidth,clip]{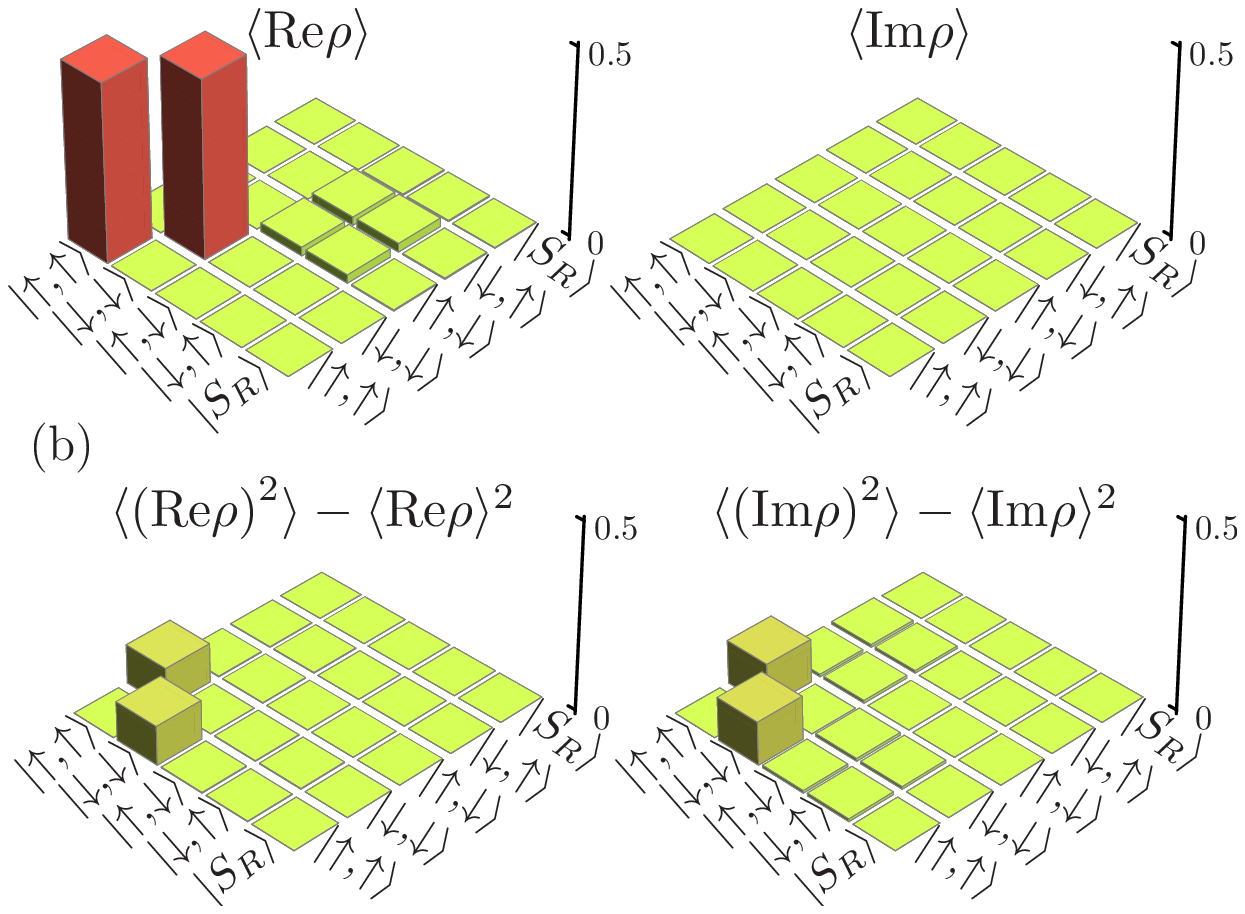}
%\end{center}
\caption{\label{pure} A pure and coherent steady state. (a) Purity of the steady state as a function of the driving frequency. At the antiresonance frequency $\Omega$ a pure state is formed. Far from resonance, the purity tends to 1/2, consistently with a spin blockaded density matrix, $\hat\rho_{SB}$.  Here, $B_x{=}B_z/5$. (b) Mean and variance of a state tomography at $\omega{=}\Omega$. Only two-electron states are represented for clarity. Finite off diagonal elements reveal a coherent superposition. Their explicit time dependence (cf. Eq.~\eqref{rst}) will be averaged out in the mean of any measurement, but will affect its variance.
}
\end{figure}

A numerical confirmation of our above results, namely that the system is driven to a pure and maximally entangled steady state, is shown next. 
The purity of the steady state is given by ${\rm Tr}(\hat\rho^2)$ which is 1 for a pure state.
%As every pure state, the dark state must be described by an idempotent density operator: $\hat\rho^2{=}\hat\rho$. Otherwise it is a mixed state. 
We verify this property for the stationary solution of the full time-dependent master equation~\eqref{mastereq}. As shown in Fig.~\ref{pure}(a), the steady state is mixed [${\rm Tr}(\hat\rho^2){<}1$, as expected for a transport configuration], but is dynamically purified at $\omega{=}\Omega$.%, ${\rm Tr}(\hat\rho^2){=}1$. 

A detection scheme for state coherence is quantum state tomography. A single-shot readout of the two electron spins can extract the required correlations~\cite{nowack-2qubit}. In our case, the explicit time dependence of the dark state off-diagonal elements, cf. Eq.~\eqref{rst}, will be averaged out in the mean of any measurement; see Fig.~\ref{pure}(b). The result would be then similar to that of a mixed spin blockade state, $\langle\hat\rho\rangle{=}(1/2)\sum_\sigma {\mid}{\sigma}{,}{\sigma}\rangle\langle{\sigma}{,}{\sigma}{\mid}$. The variance of the experimental data will discriminate between the two for the finite contribution of off-diagonal elements, cf. Fig.~\ref{pure}(b). It reveals the coherence of the steady state, in agreement with what is expected from Eq.~\eqref{rst}~\cite{timeav}.

{\it Experimental discussion---}
\label{sec:exp}
\begin{figure}[t!]
\begin{center}
\includegraphics[width=\linewidth,clip]{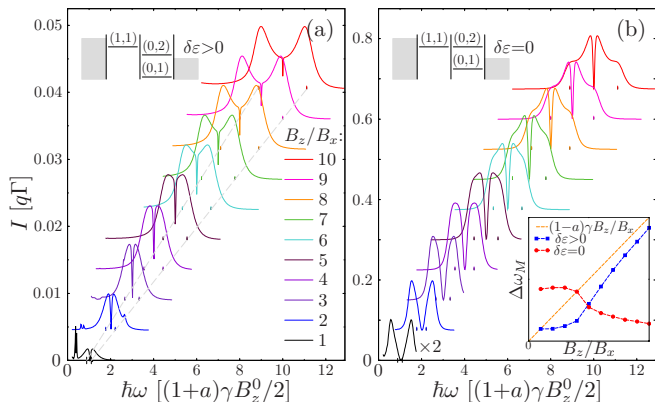}
\end{center}
\caption{\label{exp} Transport spectroscopy for different dc magnetic fields, with $B_x{=}B_z^0{=}0.1$T. (a) For finite detuning ($\delta\varepsilon{=}2$ meV), a double peak structure coinciding with the individual ESR conditions ($\hbar\omega=\Delta_l$, marked by color dots) is consistent with previous experimental observations. (b) Resonant interdot tunneling ($\delta\varepsilon{=}0$) enhances the visibility of the collective features. The dark antiresonance is robust upon increasing the magnetic field. At low $B_z$ two peaks are visible which do not correspond to individual ESR. The latter appear as humps for larger magnetic fields. An offset is introduced for clarity. Inset: Frequency splitting of the current maxima, $\Delta\omega_{M}$. In orange, the expected behaviour for the individual ESR.
}
\end{figure}
No signature of such a dark Bell state has been reported so far~\cite{nowack,pioro,nadjperge,laird}. Our results indicate that the role of coherent tunneling must be emphasized. Configurations where interdot tunneling is not resonant show well defined double peak resonances centered at the individual ESR conditions, $\hbar\omega{=}\Delta_L^z,\Delta_R^z$~\cite{nowack,pioro,nadjperge,laird}. Thus they split as $\Delta\omega_M{\approx}(1{-}a)\gamma B_z$.
Using realistic parameters, our model reproduces that behaviour for $\delta\varepsilon{>}0$, cf. Fig.~\ref{exp}(a).  Then, the effect of coherent tunneling is reduced and only a tiny antiresonance appears. 

We propose that a clearer evidence of the dark Bell state will appear in the resonant tunneling case, $\delta\varepsilon{=}0$, due to the enhanced interplay of coherent tunneling and collective spin resonance, see Fig.~\ref{exp}(b). For low $B_z$, where tunneling dominates the coherent dynamics, two peaks appear around the antiresonance condition, whose splitting weakly depends on the magnetic field, cf. inset in Fig.~\ref{exp}(b), and are not centered at $\hbar\omega{=}\Delta_L,\Delta_R$. 
At the antiresonance, current vanishes.
%There, the sharp antiresonance structure is dominant, with a  vanishing current. 
As analyzed in Ref.~\cite{michaelis}, a finite current at the dark state condition can be used to estimate the effect of other sources of decoherence different from tunneling. Upon increasing the dc magnetic field, the current develops humps that follow the individual ESR conditions and would be eventually resolved as separate peaks for larger $B_z$. However, the central structure is dominant. This peculiar behaviour, namely (i) a double peak which does not follow the individual ESR conditions at low $B_z$ and (ii) a central structure for larger $B_z$, is robust against decoherence~\cite{decoh} and conforms an unambiguous signature of the collective dark state, even if the antiresonance becomes faint. 

{\it Conclusions---}
\label{sec:conclusions}
We predict a transport-induced maximally entangled state of two spatially separated electrons in a driven double quantum dot. Interplay of coherent interdot tunneling and collective electron spin resonance is essential for leaving clear signatures of the dark Bell state in transport spectroscopy experiments. The entangled state is decoupled from its environment by parity symmetry and Pauli exclusion principle. Thus it is not affected by the decoherence due to coupling the system to leads. Other sources of decoherence can be probed and will motivate further investigation. We propose an experimental setup which is within reach where resonant interdot tunneling 
%(at the border of the spin blockade window) 
enhances the visibility of the entangled dark state features. Our work introduces a mechanism to produce Bell states in open systems for any initial condition. It will allow for investigations of non local quantum correlations of two electrons stored in solid state qubits. Our simple configuration constitutes an ideal candidate for the detection of transport dark states. 

\acknowledgments
We acknowledge support from the CSIC and FSE JAE-Doc program, the Spanish MICINN Juan de la Cierva program and MAT2011-24331, and ITN Grant No. 234970 (EU).

%%%%%%%%%%%%%%%%%%%%%%%%%%%%%%%%%%%%%%%%%%%%%%
%           APPENDIX
%%%%%%%%%%%%%%%%%%%%%%%%%%%%%%%%%%%%%%%%%%%%%%

\newpage

\begin{figure*}[h]{\Large{Dark Bell states in tunnel-coupled spin qubits --- Supplementary Information}\vspace{0.9cm}}
\begin{center}
\includegraphics[width=0.7\linewidth,clip]{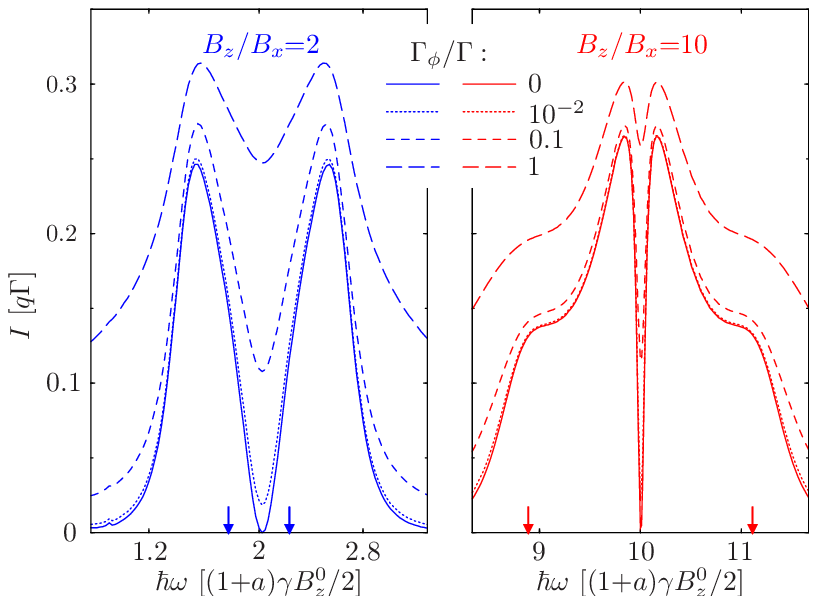}
\end{center}
\caption{Effect of decoherence as introduced in the form of a phenomenological dephasing rate. The current as a function of the driving frequency is plotted for two cases corresponding to low, $B_z/B_x{=}2$ and high dc magnetic field, $B_z/B_x{=}10$ in Fig. 4(b) of the main text. Interdot tunneling is resonant, i.e. $\delta\varepsilon{=}0$. The individual ESR conditions, $\hbar\omega{=}\Delta_L^z,\Delta_R^z$ are marked by colored arrows. The characteristic features of each case, namely {\it (i)} a double peak structure not corresponding to the individual ESR conditions at lower $B_z$ and {\it (ii)} a central structure around the antiresonance for larger $B_z$ are robust up to decoherence times of the order of the tunneling. All the parameters correspond to those of Fig. 4 in the main text, with $\Gamma^{-1}{\sim}0.6\mu$s.   }
\end{figure*}


\begin{thebibliography}{99}
\bibitem{timeav}
From a time average of Eq.~\eqref{rst} we get $\left\langle({\rm Re}\langle{\uparrow},{\uparrow}{\mid}\hat\rho_{st}{\mid}{\uparrow},{\uparrow}\rangle)^2\right\rangle=\left\langle({\rm Im}\langle{\downarrow},{\downarrow}{\mid}\hat\rho_{st}{\mid}{\downarrow},{\downarrow}\rangle)^2\right\rangle=1/4$ and $\left\langle({\rm Re}\langle{\uparrow},{\uparrow}{\mid}\hat\rho_{st}{\mid}{\downarrow},{\downarrow}\rangle)^2\right\rangle=\left\langle({\rm Im}\langle{\uparrow},{\uparrow}{\mid}\hat\rho_{st}{\mid}{\downarrow},{\downarrow}\rangle)^2\right\rangle=1/8$.
\end{thebibliography}
\end{document}